\documentclass[aip,reprint]{revtex4-1}

\usepackage{amsmath}

\usepackage{graphicx}
\usepackage{url}
\usepackage{braket}
\usepackage{paralist}
\usepackage{booktabs}
\usepackage{tabularx}
\usepackage{color}
\usepackage[capitalise]{cleveref}
\crefname{secinapp}{appendix}{appendices}
\Crefname{secinapp}{Appendix}{Appendices}

\DeclareMathOperator{\sgn}{sgn}

\begin{document}


\title{Understanding and Improving the Efficiency of Full Configuration Interaction Quantum Monte Carlo} 



\author{W. A. Vigor}
\affiliation{Department of Chemistry, Imperial College London, Exhibition Road, London, SW7 2AZ, United Kingdom}
\author{J. S. Spencer}
\affiliation{Department of Physics, Imperial College London, Exhibition Road, London, SW7 2AZ, United Kingdom}
\affiliation{Department of Materials, Imperial College London, Exhibition Road, London, SW7 2AZ, United Kingdom}
\author{M. J. Bearpark}
\affiliation{Department of Chemistry, Imperial College London, Exhibition Road, London, SW7 2AZ, United Kingdom}
\author{A. J. W. Thom}
\affiliation{Department of Chemistry, Imperial College London, Exhibition Road, London, SW7 2AZ, United Kingdom}
\affiliation{University Chemical Laboratory, Lensfield Road, Cambridge, CB2 1EW, United Kingdom}


\date{\today}

\begin{abstract}
Within Full Configuration Interaction Quantum Monte Carlo, we investigate how the statistical error behaves as a function of the parameters which control the stochastic sampling.
We define the inefficiency as a measure of the statistical error per particle sampling the space and per timestep and show there is a sizeable parameter regime where this is minimised.
We find that this inefficiency increases sublinearly with Hilbert space size and can be reduced by localising the canonical Hartree--Fock molecular orbitals, suggesting that the choice of basis impacts the method beyond that of the sign problem.

\end{abstract}

\pacs{}

\maketitle 

\section{Introduction}
The FCIQMC method\cite{BoothJCP2009,ClelandJCP2010} is a stochastic algorithm that has enabled calculation of the ground state energy of the largest molecules to date\cite{DadayJCTC2012} to FCI accuracy, i.e. exact given a fixed basis set. 
It has also found success calculating the ground state energy of model Hamiltonians such as the uniform electron gas\cite{ShepherdJCP2012} and the three band Hubbard model\cite{SchwarzPRB2015}, as well as solid state systems.\cite{BoothNature2013}
Recent developments mean that properties beyond the ground state energy, such as forces\cite{ThomasJCP2015} and excited state energies\cite{BoothJCP2012,BluntPRL2015,BluntArxiv2015} can also be calculated.
 Further, efficient stochastic implementations of other methods such as second order M\o ller Plesset Perturbation theory (MP2)\cite{ThomPRL2007,WillowJCP2012,WillowJCTC2013}, Density Functional Theory (DFT)\cite{BaerPRL2013} and Coupled Cluster theory\cite{ThomPRL2010} have been developed, forming a new field of stochastic computational chemistry.
These Monte Carlo methods differ from the more traditional Diffusion Monte Carlo method \cite{UmrigarJCP1993}, which requires the use of a fixed node approximation to prevent collapse of the wavefunction to a bosonic state. 
The eventual hope is that these methods will allow the properties and reactivity of large molecules to be investigated to unprecedented accuracy.

Quantum Monte Carlo simulations are plagued by the fermion sign problem: the quantity being sampled can be either positive or negative which may cause the calculation to converge to the wrong answer.\cite{CeperleyPRL1980,SpencerJCP2012} 
In FCIQMC the sign problem appears as the critical population of particles required to correctly sample the ground state wavefunction.\cite{SpencerJCP2012}
Whilst the scaling of this with system size has been investigated in a number of systems\cite{ClelandJCP2011,SpencerJCP2012,ShepherdPRB2014}, the effect of it on the stochastic error has not been well studied.
The (systematically improvable) initiator approximation\cite{ClelandJCP2010} can reduce the critical population by many orders of magnitude.
However, the impact of this key advance on the stochastic error has similarly not been thoroughly investigated.

The efficiency of a stochastic algorithm can be measured by computational cost (CPU time and memory requirements) necessary for convergence to a given stochastic error.
The FCIQMC algorithm is not a `black box' method and the manner in which sampling is performed can impact the computational resources used whilst performing a calculation and the resultant statistical error bar.
We believe that understanding the efficiency of a FCIQMC calculation is important for two main reasons.
Firstly so that one can use a computational budget as effectively as possible, ideally leading to the ability to perform FCIQMC calculations automatically and efficiently without careful tuning of input parameters.
Secondly it helps compare the effectiveness of different algorithms based upon and additional approximations to FCIQMC.

Previous studies of the efficiency of FCIQMC investigated the standard error as a function of computer time and focused on comparing very different algorithms.\cite{PetruzieloPRL2012,BluntJCP2015}
Here we take a different tack, running many empirical calculations to understand how the stochastic sampling of the wavefunction affects the error bar for FCIQMC in general.
Once we understand this effect, we form a metric which is independent of the parameter choices and use it to measure how the error bar scales with system size. 

We begin with an overview of the FCIQMC method. We then explore the dependence of the statistical error bar on the total population and the timestep in \cref{Eff_error_bar} and on the system size, both basis set and number of electrons, in \cref{cost-fciqmc}.
Finally, we discuss in \cref{conclusions} how the total number of particles (relative to the plateau height) affects the stochastic error bar and draw some conclusions about how to choose parameters to maximise the efficiency of the FCIQMC algorithm `a priori' from the plateau height.

\section{A Recap of the FCIQMC Method}
\label{fciqmc-recap}

The FCIQMC algorithm\cite{BoothJCP2009,ClelandJCP2010} can be viewed as a stochastic power method.
On every iteration the representation of the wavefunction, $\psi \left( \tau \right)$, at imaginary time $\tau$ is updated by sampling the action of the operator:
\begin{equation}
\psi \left( \tau + \delta \tau \right) =\left( 1 - (\hat{H} -S) \delta \tau \right)  \psi \left( \tau \right)
\label{projector}
\end{equation}
where $\hat{H}$ is the Hamiltonian, $S$ is an offset used to control normalisation, and $\delta \tau$ is the timestep.
As long as $\psi \left( 0 \right)$ has a non-zero overlap with it, the ground state will be all that remains once $\tau$ is large enough, assuming $\delta\tau$ is sufficiently small\cite{SpencerJCP2012} and $S$ is carefully controlled.
$\braket{D_i|\psi \left( \tau \right)}$ is represented by a number of (signed) unit weights located on $\ket{D_i}$; we call a single unit a psi-particle or psip.\cite{AndersonJCP1975}
Other choices of discretisation are possible\cite{UmrigarJCP1993}; for the purposes of this work we consider the simplest case.

\cref{projector} is sampled in three stages to make up a single timestep of $\delta \tau$\cite{BoothJCP2009}:
\begin{description}
\item[Spawning] each psip (with weight $w_i$) attempts to spawn a child psip on a randomly selected $\ket{D_j}$ with probability $|\braket{D_i | \hat{H} | D_j}|\delta \tau/p(j|i)$, where $p(j|i)$ is the probability that $\ket{D_j}$ is selected given the parent psip is on $\ket{D_i}$ and with sign $\sgn(-\braket{D_i | \hat{H} | D_j}w_i)$.

\item[Death] each psip dies (is removed from the simulation) with probability $|K|\delta \tau$, where $K = \braket{D_i | \hat{H} | D_i} - S$, if $K < 0$; for $K>0$, instead, a copy of the psip is made with probability $K\delta\tau$.

    \item[Annihilation] psips on the same determinant with opposite signs cancel.
\end{description}

Initially $S$ is set to the Hartree--Fock energy, which is larger (less negative) than the FCI energy and this causes the population $N_p(\tau)$ (total number of psips) to begin to grow exponentially.
If there is a sign problem\footnote{There is a sign problem in FCIQMC if it is possible that psips of opposite signs can be generated on the same determinant.\cite{SpencerJCP2012}} $N_p(\tau)$ will plateau as the sign structure is projected out.
It is only after this point that $\psi \left( \tau \right)$ becomes a stochastic representation of the wavefunction and $N_p(\tau)$ grows exponentially again.\cite{SpencerJCP2012} 
To counter this exponential growth $S$ is periodically adjusted every $A$ timesteps according to:
\begin{equation}
    S(\tau + A \delta \tau) = S(0) - \xi \log{\frac{N_p(\tau +A\delta \tau)}{N_s}},
\label{shift-1}
\end{equation}
where $S(0)$ is the initial value of the shift, $N_s$ is the population at the end of the equilibration phase and $\xi$, the damping factor, is usually fixed during a simulation.\footnote{As discussed in Ref.~\onlinecite{VigorJCP2015} this is equivalent to the original formulation of the shift in Ref.~\onlinecite{BoothJCP2009}.}

In addition to the shift, the energy of the system can be estimated via the projected estimator, which typically has a smaller statistical noise:
\begin{equation}
E_{\text{Proj}} = \frac{\braket{D_0 | \hat{H} e^{-\hat{H}\tau} | D_0}}{\braket{D_0 | e^{-\hat{H}\tau} | D_0}} =  \frac{\langle \sum_{i\ne 0} H_{0i} n_i \rangle_\tau }{\langle n_0 \rangle_\tau},
\end{equation}
where $n_i(\tau)$ is the (signed) number of psips on determinant $i$ in $\ket{\psi \left( \tau \right)}$ and $\langle \dots \rangle_\tau$ represents the time average.

The initiator approximation\cite{ClelandJCP2010} can dramatically reduce the sign problem in FCIQMC.
\cite{BoothTew_12JCP,OveryAlavi_14JCP,ThomasJCP2015,BluntPRL2015,BluntJCP2015,ShepherdPRB2012,ShepherdPRB2014,BoothNature2013}
In initiator FCIQMC (iFCIQMC), previously unoccupied determinants can only be spawned onto from determinants with a population above a certain threshold; a threshold of $3$ is typical and used throughout this paper.
The initiator approximation introduces a systematic error in the sampling of \cref{projector}, which is evident in both estimates of the energy, with unbiased sampling restored in the limit of an infinite population of psips. A monotonic convergence to the FCI energy as a function of $\langle N_p(\tau) \rangle$ has been observed for many simple systems,\cite{ClelandJCP2010} though this is not universally the case.

\section{Systems Studied}
\label{system-list}

In this paper we study the following systems: 
\begin{enumerate}
    \item An isolated neon atom in the following Dunning basis sets\cite{DunningJCP1989}: cc-pVDZ, aug-cc-pVDZ, cc-pVTZ, aug-cc-pVTZ, cc-pVQZ, aug-cc-pVQZ.
    \item The hydrogen fluoride, HF, molecule in a cc-pVDZ basis at bond lengths of $R=R_0$, $R=1.5R_0$ and $R=2R_0$, where $R_0=0.91622$\,\AA{}, the Hartree--Fock equilibrium bond length in this basis.
    \item Chains of between 2 and 7 helium atoms at intervals of $3$\,\AA{} in a 6-31G basis set.\cite{DitchfieldJCP1971}
    \item Chains of between 5 and 7 helium atoms at intervals of $3$\,\AA{} in a 6-31G basis set with localised molecular orbitals.
          We used Pipek--Mezey localisation \cite{PipekJCP1989} to localise the occupied Hartree--Fock molecular orbitals.
\end{enumerate}

Hartree--Fock calculations were performed using Q-Chem\cite{QChem}, with local modifications to obtain the required integrals. FCIQMC calculations were performed and analysed using HANDE\cite{Hande} and figures plotted using matplotlib.\cite{HunterCSE2007}  Unless otherwise stated, atomic units are used throughout.
Raw and analysed data and analysis scripts are availale at Ref.~\onlinecite{eff_data}.

\section{Population Dynamics and the Stochastic Error}
\label{Eff_error_bar}

If the Hamiltonian doesn't have a `sign problem' in FCIQMC, it is impossible for psips of opposite signs to be generated on the same determinant and so annihilation cannot occur.
In this case each psip samples the wavefunction independently from each other and hence the error in the estimate of the energy, $\sigma_E$, must behave as:
\begin{equation}
\sigma_E = \frac{a}{\sqrt{\langle N_p \rangle N \delta \tau} },
\label{decay}
\end{equation}
where $a$ is a constant of proportionality and $N$ is the number of timesteps from which $\sigma_E$ is estimated.
\cref{decay} arises because the contribution to the estimate of $E$ from each psip can be combined in a similar way to the contribution for running from multiple timesteps.
Increasing $\delta \tau$ simply increases the probability of spawning/death and so decreases $\sigma_E$ in the same way as increasing $N$ or $\langle N_p \rangle$.

$a$ can be estimated from a FCIQMC calculation using $\sigma_E$ estimated by blocking analysis of the projected energy and $\langle N_p \rangle$.
For consistency throughout this paper, we use an automatic approach to calculate the optimal block length\cite{WolffCPC2004}, which has previously been applied to DMC\cite{ConduitPRE2011} and is implemented in pyblock.\cite{pyblock}

For systems with a `sign problem', which is the usual case, $a$ will depend on $\langle N_p \rangle$ and $\delta \tau$.
We will call $a(\langle N_p \rangle, \delta \tau)$ the inefficiency and note from \cref{decay} that a smaller $a$ implies a more efficient calculation.
In Secs. \ref{scale-pop} and \ref{scal_tau} we investigate the dependence of $a$ on $\langle N_p \rangle$ and $\delta \tau$ respectively.

\subsection{Effect of population on inefficiency of FCIQMC simulations}
\label{scale-pop}

\begin{figure}[h]
\includegraphics{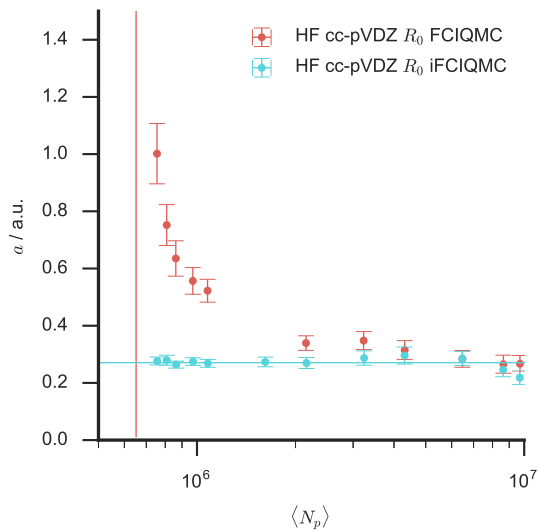}
\caption{The inefficiency $a$ as a function of the average number of psips for HF in the cc-pVDZ basis, $R_0=0.91622$\,\AA{}, and $\delta \tau=0.00175$ for both FCIQMC and iFCIQMC.
The plateau height is $\sim 6.5\times10^5$ (measured using the method described in \onlinecite{SpencerArxiv2015}) and is indicated by the vertical line.
The horizontal line shows a fit to the iFCIQMC data for constant $a$.
The initiator error is converged below the stochastic error in all iFCIQMC calculations shown; $a$ may not remain constant where this is not true.  For a population of $\langle N_p \rangle=50000$ in iFCIQMC (not shown), we find $a=0.318(18)$ compared to $a=0.2682(34)$ from the fit, showing that sufficiently small populations do have an effect on the inefficiency in iFCIQMC calculations.}
\label{pop_dynamics-np}
\end{figure}

The behaviour of $a$ with $\langle N_p \rangle$ in FCIQMC simulations is shown in \cref{pop_dynamics-np} for hydrogen fluoride.
This behaviour has been seen in a wide range of systems, which are included in the Supplemental Information.\cite{Supp}
Owing to the sign problem, only after the population reaches the plateau does the vector of psips become a stochastic representation of the eigenvector.\cite{SpencerJCP2012}
Populations below the plateau could either have a divergent $\sigma_E$ or an incorrect average correlation energy with a finite $\sigma_E$; we see the former behaviour.
For populations greater than the plateau, $a$ decays as a function of $\langle N_p \rangle$ and tends to a finite constant in the large population limit.

If the initiator approximation is used, then we find that $a$ is a constant as a function of $\langle N_p \rangle$ for populations much smaller than the plateau and, for a fixed $\delta \tau$, the same as $a$ in FCIQMC in the large $\langle N_p \rangle$ limit.
We call this limit on $a$ the iFCIQMC limit.
For example in \cref{pop_dynamics-np} inefficiency in FCIQMC doesn't hit the iFCIQMC limit until about $\langle N_p \rangle = 2\times10^6$, approximately $3$ times the plateau height.
This is an important point: if the largest population affordable in FCIQMC is not sufficient to reach the iFCIQMC limit (but is sufficient to exceed the plateau), then the initiator approximation is still useful as it provides a significant reduction in the stochastic error for the same computational cost.
It may, however, be difficult to quantify if the introduction of an initiator error is a price worth paying for a potentially significant reduction in statistical error bar.

\begin{figure*}[t]
\includegraphics{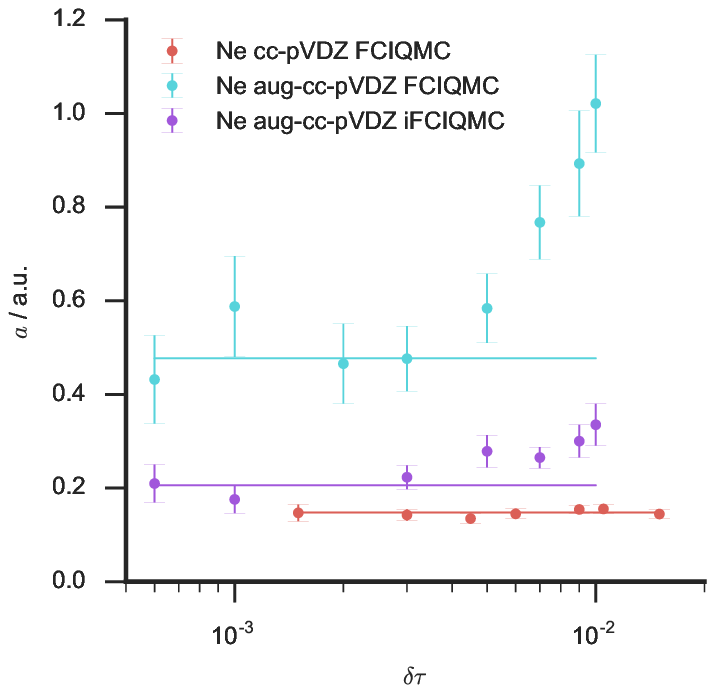}
\includegraphics{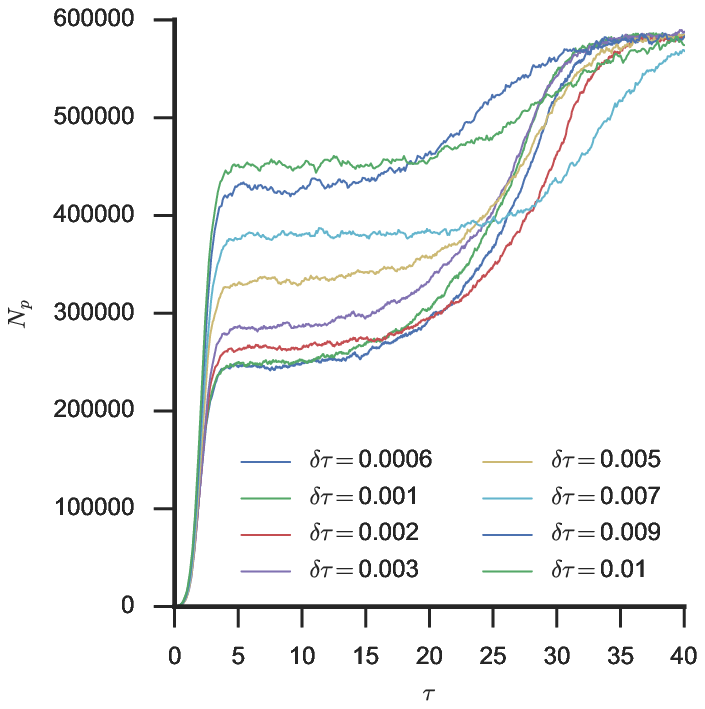}
\caption{Left: The inefficiency $a$ as a function of $\delta \tau$ for Ne in cc-pVDZ ($\langle N_p \rangle \approx 46100$) and aug-cc-pVDZ bases ($\langle N_p \rangle \approx 590000$).
Once $\delta \tau$ is large enough to raise the plateau height above $\langle N_p \rangle$, $a$ diverges in a similar way as in \cref{scale-pop}.
Right: The plateau height as a function of the timestep for FCIQMC on Ne in aug-cc-pVDZ; note the region in the plateau is roughly minimum matches the similar region for $a$.
}
\label{pop_dynamics-tau}
\end{figure*}

This behaviour has implications about how best to run parallel FCIQMC and iFCIQMC simulations given a fixed amount of computational resources.

In the canonical parallel FCIQMC implementation\cite{BoothMP2013} the Hilbert space is partitioned over the processors, resulting in efficient distribution of the memory demands across the processors.
Psips spawned from a parent determinant in one part of the Hilbert space onto a child determinant in another part of the space are communicated between processors.
It is this communication overhead that limits the parallel scalability of FCIQMC.

The simplest way to parallelise a Monte-Carlo algorithm is to run independent calculations and combine statistics gathered in each calculation.
The overall stochastic error scales as $1/\sqrt{N_I}$, where $N_I$ is the number of independent calculations.
Therefore once the population is such that $a$ reaches the iFCIQMC limit, it is more efficient to use multiple independent simulations to minimise interprocess communication.

\subsection{Effect of timestep on inefficiency of FCIQMC simulations}
\label{scal_tau}

$a$ is approximately constant for a sufficiently small timestep, $\delta \tau$, and otherwise $a$ increases with $\delta \tau$;
the value of $\delta \tau$ after which $a$ is non-constant is system dependent.
We have seen this behaviour for all systems investigated in this paper (see Supplemental Information\cite{Supp}) and for FCIQMC and iFCIQMC.
\cref{pop_dynamics-tau} shows this for an isolated neon atom in cc-pVDZ and aug-cc-pVDZ bases.

A previous investigation \cite{booth2010novel} has shown that the plateau height exhibits a similar behaviour to $a$ as a function of $\delta \tau$ in FCIQMC calculations.
As shown in \cref{pop_dynamics-tau} for the Ne atom (and in the Supplemental Information\cite{Supp} for other systems) the plateau height appears to be a good metric for when $a$ remains constant as a function of $\delta \tau$.
This is useful as the plateau height is easier and cheaper to measure than statistical accumulation of the energy and hence evaluation of $a$.
We denote the largest $\delta \tau$ such that $a$ is constant as $\delta \tau_0$.

In the Supplemental Information\cite{Supp} we investigate whether $\langle N_p \rangle$ has an impact on $\delta \tau_0$ and find this is not the case.
We do however find that $a$ increases faster as a function of $\delta \tau$ when $ \delta \tau > \delta \tau_0$ for a small $\langle N_p \rangle$ than a large $\langle N_p \rangle$ and hence the most efficient timestep has an implicit dependency on $\langle N_p \rangle$.

The population dynamics can become unstable if the $\delta \tau$ is set large enough such that the exponential growth cannot be countered by population control.
We find that the point at which the population explodes is beyond $\delta \tau_0$ for the systems studied here with the exception of Ne cc-pVDZ (see \cref{pop_dynamics-tau} left) which doesn't have a noticeable plateau.\footnote{When running the calculations for \cref{pop_dynamics-tau} top right we found that the population explodes when $\delta \tau$ is slightly greater than 0.05 for Ne cc-pVDZ.}

\section{The scaling of inefficiency as a function of system size}
\label{cost-fciqmc}

\begin{figure}[h]
\includegraphics{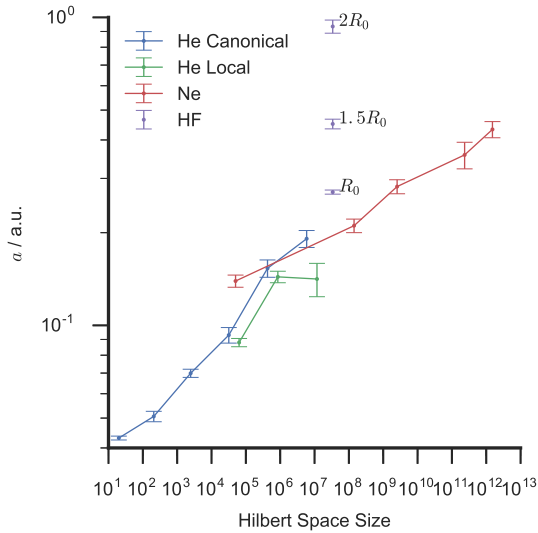}
\caption{The scaling of $a_{\text{min}}$ as a function of system size.
\cref{system-list} describes the systems studied in more detail.
$a_{\text{min}}$ was calculated by fitting $a$ for iFCIQMC calculations with differing populations with the exception of Ne cc-pVDZ, He2, He3 and He4, which have no noticeable plateau, where many FCIQMC calculations were used.
The timestep was set to be sufficiently small such that $a$ is constant.
The Supplemental Information\cite{Supp} shows the studies used to find $a_{\text{min}}$ for each system.
}
\label{a-scales}
\end{figure}

In the previous sections we investigated both the scaling of the inefficiency, $a$, as a function of (mean) population, $\langle N_p \rangle$, and timestep, $\delta \tau$.
Following this, we can define a metric that lets us investigate the scaling of the error bar as we change system.
The simplest such metric is to find the minimum value of $a$, $a_
\text{min}$, i.e. when it is a constant as a function of $\langle N_p \rangle$ and $\delta \tau$.
This requires calculations with $\delta \tau<\delta \tau_0$ and either the initiator approximation (with the initiator error converged) or a large $\langle N_p \rangle$.
The behaviour of $a_{\text{min}}$ as a function of Hilbert space size is shown in \cref{a-scales} for all systems studied in this work: the neon atom, hydrogen fluoride at different bond lengths and chains of helium atoms in the canonical and localised basis sets.
We see a sublinear relationship between the scaling of $a_{\text{min}}$ with size of Hilbert space for the same chemical species and a significant impact when the degree of strong correlation is increased.

FCIQMC is very efficient at finding the ground state of the neon atom: $f_c$, the ratio of plateau height to Hilbert space size, is $\sim 10^{-4}$.\cite{BoothJCP2009}
The size of the Hilbert space scales factorially with the basis set yet there is a sublinear scaling of the stochastic error with the Hilbert space size.

In contrast $f_c$ scales superlinearly for chains of helium atoms separated by $3$ \AA{} in the canonical Hartree--Fock basis: for He$_4$ $f_c=0.3$ whereas for He$_7$ $f_c=0.97$.
However, $a_{\text{min}}$ again seems to scale sublinearly with the size of the Hilbert space.
The similar behaviour, within statistical error bars, of $a_{\text{min}}$ for Ne atom and helium chains suggests sublinear scaling of the stochastic error in FCIQMC with Hilbert space size for weakly correlated systems with no strong dependence on the FCIQMC sign problem.
In a similar style study in DMC\cite{NemecPRB2010}, Nemec et al. investigated computational scaling as a function of the number of hydrogen atoms at a large separation, finding that the computer time required to achieve a fixed error bar scales exponentially as the square root of the system size.
As the Hilbert space size scales loosely exponentially with the system size, we see this trend is similar, though with such few systems studied it seems premature to generalize this trend.

Localisation of the molecular oribitals breaks symmetry and so increases the size of the accessible Hilbert space.
However we find in this case that the plateau height decreases\footnote{Note that this behaviour is system specific.  Ref.~\onlinecite{booth2010novel} found that systems where the plateau height was unaffected by orbital localisation.};
for example localisation causes the plateau for He$_7$ to decrease from $5.7 \times 10^6$ ($f_c=0.97$) to $4.6 \times 10^6$ ($f_c=0.39$).
Importantly localisation also appears to decrease $a_{\text{min}}$ compared to using the canonical orbitals, especially for He$_7$, though it is difficult to draw definitive conclusions from three data points with non-negligible standard errors.

The behaviour of $a_{\text{min}}$ is not solely governed by the size of the Hilbert space, however.
Stretching hydrogen fluoride in a fixed basis set (and hence Hilbert space) causes the correlation energy to increase.
We find that $a_{\text{min}}$ increases significantly with bond length but have not found a simple linear scaling with the correlation energy or plateau height.
Given this behaviour doesn't fit with the weakly correlated systems studied above, investigation of the scaling of the error bar in FCIQMC in the strongly correlated limit would be an interesting topic for future investigation.

\section{Conclusions}
\label{conclusions}

We have defined a metric, {\em inefficiency}, for measuring the stochastic efficiency of FCIQMC calculations and investigated its behaviour as a function of the parameters controlling the population dynamics and as a function of system size.
We have found that a sizeable reduction in the stochastic error is possible by increasing the population of psips and using the largest possible timestep such that the plateau height remains constant.\footnote{This can be easily assessed prior to production calculations given that propagation to the plateau is computationally cheap compared to an entire FCIQMC calculation and the plateau height can easily be found in an automated fashion.\cite{ShepherdPRB2014,SpencerArxiv2015}}
The optimal timestep is transferable between FCIQMC and iFCIQMC simulations.
The efficiency decreases (and inefficiency increases) sublinearly with the size of the Hilbert space in weakly correlated systems.
The population required to exceed the plateau height or converge the initiator approximation becomes a limiting factor much faster than the rise in the stochastic error bar for a given computational effort.
This suggests that if improved approximations which reduce the sign problem can be made, then the stochastic noise will not be insurmountable when treating yet larger systems.

This analysis ignores the impact of population and timestep on the computational cost of the calculation, which is detailed in \Cref{comp-cost}.
Ideally we would like to set the population and timestep such that the stochastic error decays as fast as possible as a function of the computational effort.
This analysis doesn't change the guidelines given above: 
\begin{inparaenum}[i)]
    \item the population should be large enough such that the inefficiency metric, $a$, is converged to the iFCIQMC limit;
    \item the largest timestep for which the plateau remains constant represents a lower bound on the most efficient timestep and is difficult to improve upon without running lots of expensive calculations at different timesteps.
\end{inparaenum}

Orbital localisation was found to be effective for reducing the plateau height, and thus the cost of a FCIQMC calculation for chains of helium atoms, as well improving the stochastic efficiency.
Consequently, it may be worthwhile investigating FCIQMC algorithms tuned to localised orbitals.

More broadly, we have provided a framework for assessing the statistical efficiency of FCIQMC and related methods.
Given the recent activity in such methodological development\cite{ThomPRL2010,PetruzieloPRL2012,BluntPRB2014,MaloneJCP2015,TennoJCP2013,KolodrubetzPRB2012}, this approach offers a useful means to fairly compare alternative implementations and algorithms.

\begin{acknowledgments}
We are indebted to Dr. George Booth, Prof. Ali Alavi and Prof. Matthew Foulkes for enlightening discussions.
WAV is grateful to EPSRC for a studentship and for a knowledge transfer secondment under Grant No.~EP/K503733/1.
JSS acknowledges the research environment provided by the Thomas Young Centre under Grant No.~TYC-101.  
AJWT thanks Imperial College for a Junior Research Fellowship and the Royal Society for a University Research Fellowship under Grant No.~UF110161.
Calculations were performed using the Imperial College High Performance Computing Service.\cite{ICHPC}   
\end{acknowledgments}

\appendix
\section{Computational cost of an FCIQMC calculation}
\label[secinapp]{comp-cost}
\begin{figure}[h]
\includegraphics{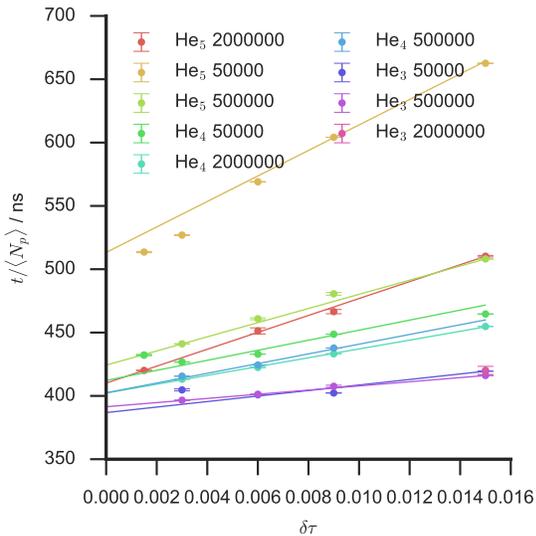}
\caption{The computer time of a single iteration per psip as a function of the timestep.
There is a linear relationship (\cref{timing-fciqmc}) with the timestep and a near-linear relationship with the number of psips for a large population.
All calculations were run on a single core of an Intel Core i7-2600 processor.}
\label{cost-per-step}
\end{figure}

An efficient FCIQMC implementation uses a sparse storage scheme to store the list of psips: a representation of a determinant along with the number of psips occupying the determinant are stored together.\cite{BoothMP2013}
The expensive steps are evolving the psips (spawning and death) and annihilation (combining psips on the same determinant).

The cost of spawning and death is linear in the population; the death step can be efficiently performed on a per-determinant rather than per-psip basis but the effect of this can absorbed into a system-dependent prefactor.  The prefactor also includes the cost of selecting a random excitation and calculating the necessary Hamiltonian matrix elements.

The cost of annihilation depends on the number of psips created in a single timestep.  Assuming the calculation has equilibrated at a population in excess of the critical population, the number of psips created scales linearly with the population and the timestep with a rate dependent on the average absolute value of the off-diagonal elements in the Hamiltonian matrix.  Whilst annihilation can be performed with a linear scaling\cite{BoothMP2013}, the implementation in HANDE currently uses a binary search approach that introduces a logarithmic dependency on the number of occupied determinants.  For the purposes of this analysis, we shall neglect the logarithmic dependency and instead focus on the optimal case.

Under these assumptions the average total amount of computer time to perform a single timestep $\langle t \rangle$ is:
\begin{equation}
\langle t  \rangle = C_1 \langle  N_p \rangle  + C_2 \delta \tau \langle N_p \rangle.
\label{timing-fciqmc}
\end{equation}
$C_1$ and $C_2$ depend on both the computer architecture and the chemical system of interest.
All timing calculations were run on one processor (core) of the same machine.\footnote{By running on a single core we avoid complications resulting from $ N_p $ affecting load balancing, which will in turn affect $t$.
Load balancing has been extensively investigated in Ref.~\onlinecite{BoothMP2013}.}

In \cref{cost-per-step} we show the computational time, $t$, of a single timestep per psip for a range of different populations and systems.
If \cref{timing-fciqmc} was obeyed exactly then there to be no dependence between the number of psips and the intercept or gradient of \cref{cost-per-step} for a given system; the approximations made are good at large $\langle N_p \rangle$.  The approximation is less suitable for a small population as it assumes that the population is such that the average number of occupied determinants is approximately constant.

From \cref{timing-fciqmc} and \cref{decay}, it follows that
\begin{equation}
\sigma_E = \frac{a(\langle N_p \rangle, \delta \tau) \sqrt{C_1/{\delta \tau} + C_2 }}{\sqrt{N \langle t \rangle}}.
\end{equation}
As $N \langle t \rangle$ is the total computer time for running for $N$ steps, minimising $a^\prime = a \sqrt{C_1/{\delta \tau} + C_2 }$ gives the most efficient population dynamics.
Whilst the exact form of $a(\langle N_p \rangle, \delta \tau)$ is not known, we can draw qualitative conclusions from the behaviour observed in \ref{scale-pop} and \ref{scal_tau}:
\begin{inparaenum}[i)]
\item $a^\prime$ and $a$ share the same dependence on $\langle N_p \rangle$; FCIQMC calculations should therefore be performed with sufficiently large population to place it in the iFCIQMC limit;
\item if $\delta \tau< \delta \tau_0$ then $a^\prime$ will decrease as $\delta \tau^{-1/2}$, given that $a$ is constant with $\delta \tau$ in this region.
If $\delta \tau > \delta \tau_0$ $a$ will increases with $\delta \tau$ which will compete with (and dominate) the $\sqrt{C_1/{\delta \tau} + C_2 }$ term, making it difficult to quantify what will happen to $a^\prime$.
Thus $\delta \tau_0$ forms a lower bound on the most efficient timestep.
\end{inparaenum}



%
%

%



%

\end{document}